\documentclass[runningheads]{llncs}
\usepackage[T1]{fontenc}
\usepackage{microtype}
\usepackage{graphicx}
\usepackage{booktabs}
\usepackage{url}

\begin{document}
\title{Determinants of Starting Salaries for Filipino Graduates: An Explainable Machine Learning Approach}
\titlerunning{Determinants of Starting Salaries for Filipino Graduates}
\author{
    Alexander Gabriel A. Aranes\inst{1,2} \and
    John Michael C. Magpantay\inst{1,2} \and
    Reginald Neil C. Recario\inst{1,2} \and
    Jamlech Iram N. Gojo Cruz\inst{1,2} \and
    Rodolfo C. Camaclang III\inst{1,2}
    }
\authorrunning{A. Aranes et al.}
\institute{
Institute of Computer Science, University of the Philippines Los Baños, Philippines \and
Machine Learning and Artificial Intelligence Applications Lab, University of the Philippines Los Baños, Philippines\\
\email{\{aaaranes, jcmagpantay, rcrecario, jngojocruz, rccamaclang1\}@up.edu.ph} 
}
\maketitle
\begin{abstract}
Filipino graduates face a persistent disconnect between educational preparation and labor market outcomes, where starting salary is a key signal of entry-level valuation. Current Philippine research is dominated by descriptive tracer studies that document employment rates but do not explain the determinants of pay. We address this gap using a crowd-sourced survey dataset of graduate responses whose noisy, self-reported nature makes it a challenging prediction target. Applying machine learning to this problem, we identify job role and industry as the dominant determinants of starting salary, significantly outweighing institutional prestige. The strength of this finding is its central contribution: it is corroborated by three independent lines of evidence, namely SHAP attributions, the heavy reliance of the best ensemble on occupational text, and a Natural Language Inference reformulation. These results suggest that career guidance and policy should prioritize sector-specific skills over institutional brand.

\end{abstract}

\section{Introduction}

The continuous entry of new graduates into the workforce has intensified scrutiny of the alignment between education and labor market outcomes. In the Philippines, this misalignment is acute: college graduates comprised 25.6\% of the total unemployed adult labor force in 2023 \cite{lu2024}. This structural mismatch often manifests as a `horizontal mismatch,' where graduates prioritize sectors with higher wage premiums, such as Business Process Outsourcing (BPO), over degree-aligned careers \cite{bender2011,paqueo2016}. Among employment quality metrics, starting salary is uniquely informative; it serves as a direct market signal of how much the labor market values a graduate’s background at the point of entry. Understanding the drivers of this valuation is critical not only for graduates but for the educators and policymakers who shape the professional pipeline \cite{wang2022}.

Graduate outcomes in the Philippines are traditionally monitored through university tracer studies \cite{albina2020tracer,tutor2019gts}. While valuable for describing employment rates and average pay, these studies are primarily descriptive; they document what happened but do not explain why some graduates earn more than others, nor do they predict outcomes for incoming cohorts. The gap between descriptive reporting and predictive, explanatory analytics remains a significant barrier to evidence-based career guidance and policy formulation.

Addressing this gap, this paper applies machine learning to graduate salary prediction in the Philippines, establishing a predictive baseline and systematically investigating the design choices that constrain it. Classical machine learning classifiers, interpreted through Shapley Additive Explanations (SHAP) analysis, identify the most influential salary determinants. The baseline investigation reveals a performance ceiling that prompted a systematic inquiry into its source, examining whether the culprit was the feature representation, the model architecture, or the task design itself.

Specifically, this paper makes three core contributions:
\begin{enumerate}
    \item An XAI-grounded finding that job role and industry are the dominant determinants of starting salary in the Philippine labor market, significantly outweighing institutional prestige. This is corroborated by three independent lines of evidence: SHAP attributions, ensemble weighting, and NLI reasoning.
    \item Two domain-specific methodological lessons for ML-based salary classification, surfaced through exploratory extensions: raw occupational text (via TF-IDF and transformer representations) outperforms standard occupational classification codes, and economically motivated label boundaries outperform statistical quartiles. These lessons are transferable to future research in this and analogous labor markets.
    \item A characterization of the dataset's predictive ceiling, showing that the limited and noisy crowd-sourced feature set, rather than model capacity, is the binding constraint, and that gains from task reformulation come at the cost of predictive resolution.
\end{enumerate}

\subsection{Background and Related Work}
Classical machine learning classifiers, particularly tree-based ensembles such as Random Forest and XGBoost, have dominated salary prediction tasks due to their interpretability and performance in structured, low-resource settings \cite{henshaw2025,martin2018salary}. Recent work extends these with Natural Language Processing, where reformulating classification as a Natural Language Inference (NLI) task, that is, scoring whether a text premise entails a class hypothesis, enables effective prediction in low-resource domains \cite{yin-etal-2019-benchmarking,hegselmann2023tabllm}. Section~\ref{subsec:extensions} details our use of TF-IDF representations, transformer fine-tuning, and an entailment-based NLI reformulation as exploratory extensions to the classical baseline.

The graduate salary literature identifies three broad determinant pillars: human capital (academic performance, degree), social capital (university prestige, networks), and labor market signals (job role, industry) \cite{wang2022,henshaw2025,Basir2023CHAID}. Philippine work in this area has consisted mainly of descriptive tracer studies \cite{dzomeku2024tracer,albina2020tracer,bautista2023tracer}. This study bridges that gap by applying predictive, explainable modeling to a localized, crowd-sourced graduate dataset.

\section{Methods}

\begin{figure}[ht]
    \centering
    \includegraphics[width=\textwidth]{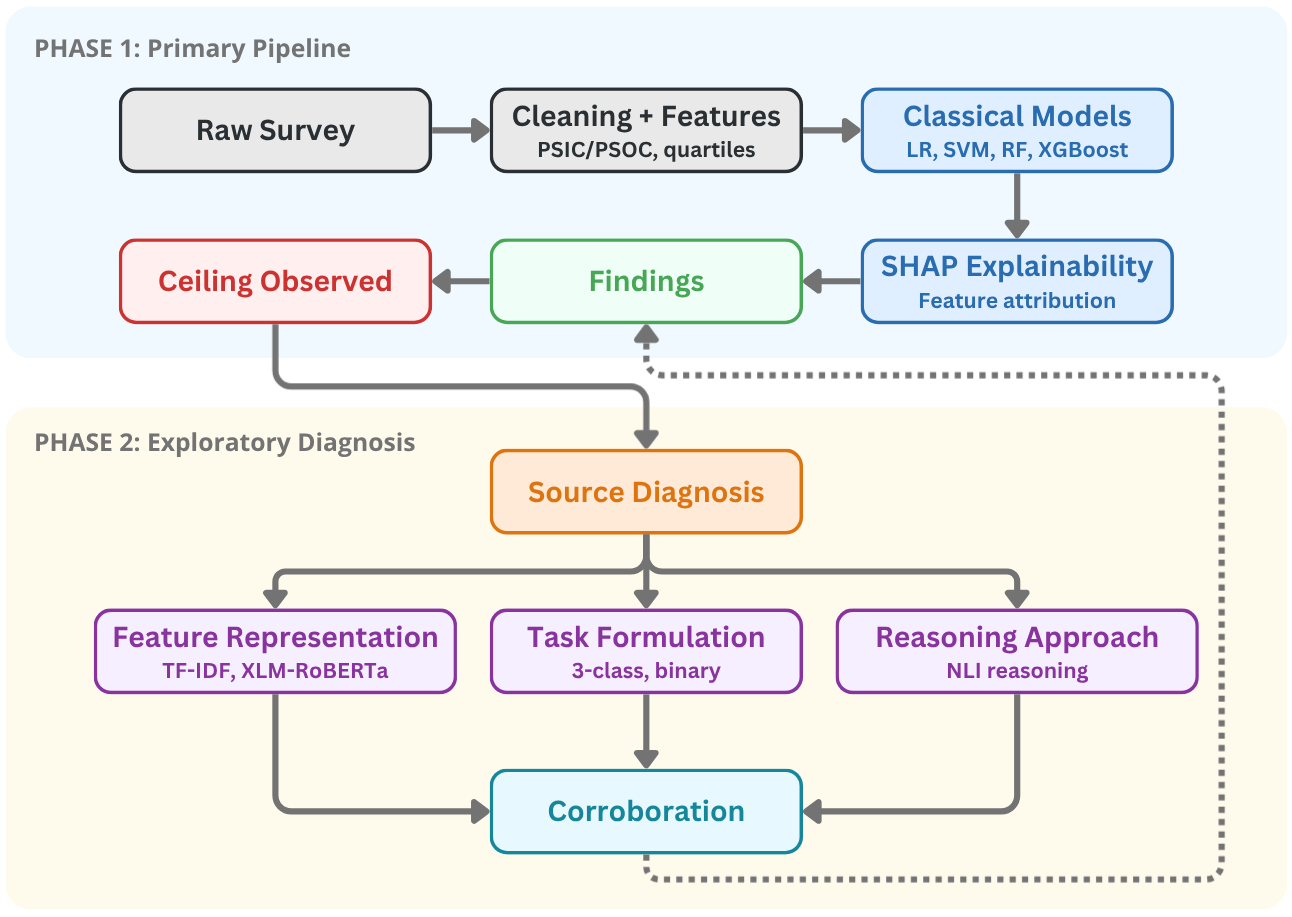}
    \caption{Two-phase methodology. Phase 1 (top) is the primary pipeline: a classical baseline with SHAP explainability that yields the central finding and reveals a performance ceiling. Phase 2 (bottom) diagnoses the source of the ceiling along three tracks (feature representation, task formulation, and reasoning approach), whose results corroborate the original finding (dashed feedback loop).}
    \label{fig:pfd}
\end{figure}

The methodological framework of this study follows a structured analytical pipeline that transitions from raw survey data to interpretable predictive insights. As shown in Figure \ref{fig:pfd}, the work proceeds in two phases: a primary pipeline that produces the explainable finding and reveals a performance ceiling, and an exploratory phase that diagnoses the source of that ceiling and corroborates the finding.

\subsection{Dataset}
This study analyzes data from the ``First Pay'' salary survey, an initiative conducted by the EdTech startup ``Liyab'' in the Philippines. The dataset represents a unique, crowd-sourced collection of initial employment outcomes from Filipino graduates across various higher education institutions (HEIs). By leveraging social media channels for distribution from 2020 through June 2025, the survey captured a broad demographic of early-career professionals, originally garnering 2,934 raw responses.
 
Because the data are self-reported through an open online form rather than a controlled instrument, they are inherently noisy: job roles and industries are free-text entries with inconsistent spelling, abbreviation, and granularity (e.g., ``SWE,'' ``software engineer,'' and ``software dev'' all appear); salaries cluster at round numbers; and several determinants known to influence pay, such as company size, geographic region, and years of experience, are absent entirely. Table~\ref{tab:dataset_schema} summarizes the available fields and their types. This combination of free-text predictors, missing confounders, and self-selection makes the dataset a genuinely challenging prediction target, and as the results show, it is central to understanding the performance ceiling reported in Section~\ref{sec:results}.

\begin{table}[t]
\centering
\caption{Schema of the ``First Pay'' survey dataset.}
\label{tab:dataset_schema}
\begin{tabular}{lll}
\toprule
\textbf{Field} & \textbf{Type} & \textbf{Role in study} \\
\midrule
Monthly Salary (PHP) & Numeric (continuous) & Target variable (discretized) \\
Year Started & Numeric (ordinal) & Predictor \\
Industry & Free text (categorical) & Predictor (primary signal) \\
Job Role & Free text (categorical) & Predictor (primary signal) \\
University / HEI & Free text (categorical) & Predictor \\
Gender & Categorical (3 levels) & Predictor \\
Negotiated Salary & Boolean (Yes/No) & Predictor \\
\bottomrule
\end{tabular}
\end{table}

\subsection{Data Preprocessing}
The initial phase of data cleaning focused on standardizing raw entries to eliminate noise. To ensure chronological relevance and remove erroneous data, records were filtered to include only those with graduation dates between 1987 and 2025. Following this, categorical consistency was addressed by mapping gender entries into three distinct categories (Male, Female, and Others) to rectify misspellings and inconsistent formatting. To further refine the numerical integrity of the dataset, the influence of extreme salary outliers was mitigated using the Interquartile Range (IQR) method, ensuring that skewed values did not negatively affect model training.
    
Building upon the cleaned data, feature engineering was employed to enhance analytical depth. A primary transformation involved discretizing quantitative salary values into four quartile-based income brackets (\textit{Low, Lower-Mid, Upper-Mid,} and \textit{High}) to facilitate a more structured comparative analysis \cite{martin2018salary}. Similarly, university classifications were simplified into two institutional types, Private and State, to streamline the modeling process. To maintain longitudinal accuracy, quantitative salary figures were approximately adjusted for inflation using historical Consumer Price Index (CPI) values from the Philippine Statistics Authority \cite{PSA_CPI}, converting reported salaries to approximate constant-price equivalents so that entries spanning several decades are comparable in real terms. Finally, to address the high variability and lack of standardization in raw responses, job industries and roles were consolidated by mapping them to the Philippine Standard Industrial Classification (PSIC) codes \cite{PSA_PSIC_2019} and the Philippine Standard Occupational Classification (PSOC) codes \cite{PSA_PSOC_2022} respectively.

\subsection{Baseline Modeling and Evaluation}
The dataset was stratified into an 80\% training and 20\% testing split. To assess the impact of feature representation, we compared two encoding pipelines: sparse One-Hot Encoding (OHE) and dense, target-based CatBoost encoding \cite{Hancock2020}. These were tested across four classifier architectures (Logistic Regression, SVM, Random Forest, and XGBoost) that span linear, geometric, and non-linear decision boundaries. Hyperparameters were tuned via randomized search with stratified 5-fold cross-validation in the scikit-learn ecosystem.

Performance was evaluated on the held-out test set using the Weighted F1-score as the primary metric. Because the quartile classes are only mildly imbalanced (23.9--25.9\%), weighted F1 accounts for class support without being biased by slight majority classes; we also report weighted precision and recall. For explainability, we applied SHAP \cite{lundberg2017shap} to the best-performing model, following prior work that uses feature attribution to interpret salary-prediction models \cite{wang2022,henshaw2025}. Rooted in game theory, SHAP quantifies each feature's contribution to the prediction, allowing us to rank determinants such as job role or university type and move beyond ``black box'' outputs. This forms the basis for the study's central explainable finding.

\subsection{Exploratory Extensions}
\label{subsec:extensions}
The accuracy ceiling of the classical baseline (reported in Section~\ref{sec:results}) prompted a set of exploratory experiments to isolate \textit{why} it arises, that is, whether the cause is the feature representation, the model architecture, or the task design. Our two initial design choices (PSIC/PSOC occupational codes and quartile-based labels) were reasonable starting hypotheses, each examined in turn. For this phase, stricter cleaning (approximate CPI inflation adjustment anchored to a 2018 reference year; occupational title normalisation) refined the dataset to 2,763 valid observations, which we partitioned into an 80/10/10 train/validation/test split (stratified on the original four-class label).
 
\textbf{Feature representation.} We first replaced the PSIC/PSOC codes, which may collapse salary-predictive textual detail into coarse categories, with TF-IDF unigram and bigram features extracted directly from the raw, concatenated job role and industry text (sublinear scaling, minimum document frequency of 2). Bigrams preserve role-specific expressions such as ``software engineer,'' and Logistic Regression was retained as the classifier for continuity with the baseline. We then fine-tuned a 270M-parameter multilingual transformer, \textit{xlm-roberta-base}, whose contextual representations capture similarity between lexically distinct but semantically related roles (e.g., ``data scientist'' and ``machine learning engineer''). The transformer was trained for 15 epochs (learning rate $2\times10^{-5}$, batch size 32) with a Focal Loss objective ($\gamma=2.0$) that emphasizes hard near-boundary records. Finally, the class probabilities of LR+TF-IDF and the fine-tuned transformer were combined by weighted soft voting. The transformer mixing weight $\alpha$ was tuned on the validation set described above.

\textbf{Task and reasoning reformulation.} To test whether the ceiling stemmed from the quartile labels, we introduced two economically grounded alternatives: a \textbf{3-class economic} split (Low $<$ PHP 16,000; Mid PHP 16,000--24,000; High $>$ PHP 24,000), aligned to real minimum-wage and market tiers, and a \textbf{binary median split} at PHP 18,500. We further reformulated classification as a Natural Language Inference (NLI) task: each record becomes a natural-language premise describing the graduate, paired with one economic-class hypothesis (e.g., ``This person earns a market-rate salary between PHP 16,000 and PHP 24,000 per month''). Rather than using an off-the-shelf entailment model, we fine-tuned \textit{xlm-roberta-base} as a binary entailment classifier and assigned each record the highest-scoring hypothesis. This lets the model reason over the economic \textit{meaning} of a salary band rather than keyword frequencies alone. For the binary task, a multi-boundary NLI ensemble combined entailment models trained at three thresholds (PHP 16,000, 18,500, and 24,000).

\section{Results and Discussion}
\label{sec:results}

\begin{figure}[t]
    \centering
    \includegraphics[width=\linewidth]{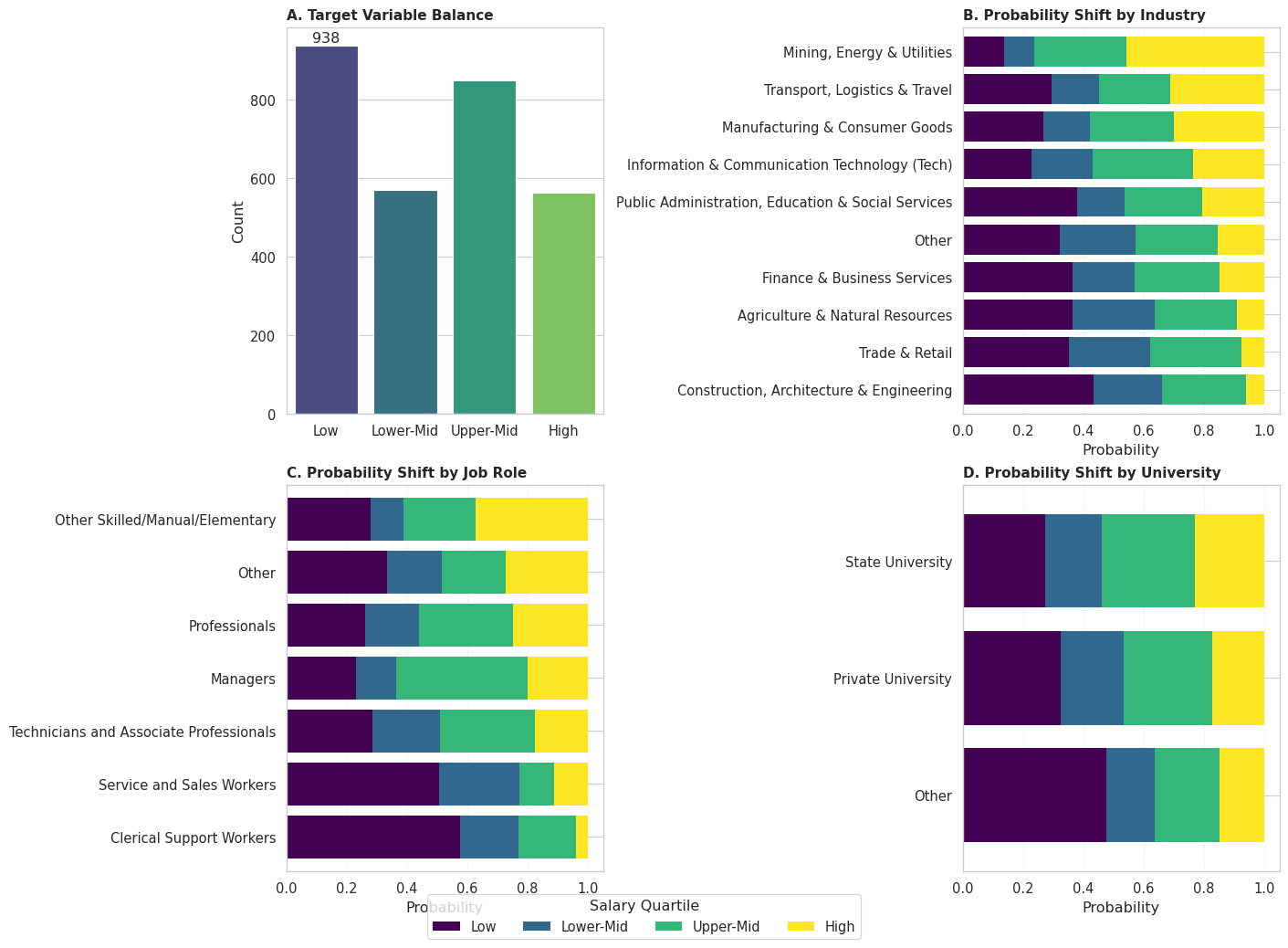}
    \caption{Exploratory data analysis grid, showing (a) balanced salary quartiles, (b and c) distinct probability shifts by Industry and Role confirming signal separability, and (d) similar distributions across university types indicating weak predictive differentiation.}
    \label{fig:eda_grid}
\end{figure}

\subsection{Dataset Characteristics}
After cleaning, the baseline analysis set contains 2,922 observations distributed across four salary quartiles (Low 25.3\%, Lower-Mid 25.9\%, Upper-Mid 23.9\%, High 25.0\%), with 50.4\% female respondents, 49.2\% from private universities, and only 14.5\% who negotiated their starting salary; the stricter cleaning of the exploratory phase (Section~\ref{subsec:extensions}) further reduced this to 2,763 observations. Figure~\ref{fig:eda_grid} confirms clear signal separability: Industry and Job Role exhibit distinct salary probability shifts, where high-income bands expand markedly in Mining and Managers compared to Construction and Clerical. In contrast, University Type distributions are similar across quartiles, indicating weak predictive differentiation.

\begin{table}[ht!]
\centering
\caption{Model performance comparison of One-Hot vs. CatBoost encoding. Values in bold indicate the best performance for each metric within this comparison.}
\label{tab:model_performance}
\renewcommand{\arraystretch}{1.2}
\begin{tabular*}{\textwidth}{@{\extracolsep{\fill}}llcccc}
\toprule
\textbf{Encoding} & \textbf{Model} & \textbf{Accuracy} & \textbf{F1-Score} & \textbf{Precision} & \textbf{Recall} \\
\midrule
One-Hot  & Logistic Regression & \textbf{0.431} & 0.390 & \textbf{0.397} & \textbf{0.431} \\
         & XGBoost             & 0.427 & \textbf{0.397} & 0.392 & 0.427 \\
         & SVM                 & 0.426 & 0.386 & 0.386 & 0.426 \\
         & Random Forest       & 0.424 & 0.382 & 0.367 & 0.424 \\
\midrule
CatBoost & XGBoost             & 0.393 & 0.351 & 0.369 & 0.393 \\
         & Logistic Regression & 0.391 & 0.327 & 0.341 & 0.391 \\
         & Random Forest       & 0.391 & 0.338 & 0.310 & 0.391 \\
         & SVM                 & 0.388 & 0.311 & 0.385 & 0.388 \\
\bottomrule
\end{tabular*}
\end{table}

\subsection{Model Performance}

Table \ref{tab:model_performance} presents the ablation between sparse (OHE) and dense (CatBoost) feature representations. One-Hot Encoding consistently outperformed CatBoost across all four classifier architectures, yielding an average Weighted F1-score improvement of approximately 4.6\%. The performance degradation in the dense representation was particularly severe for the linear models, where the Weighted F1-score dropped from 0.390 (OHE) to 0.327 (CatBoost).

Logistic Regression with OHE achieved the highest overall accuracy (43.1\%, F1w: 0.390), matching the more complex XGBoost (F1w: 0.397), suggesting the salary-determinant relationship is predominantly linear. CatBoost encoding caused a ``sensitivity collapse'' in the High salary quartile, with a 67\% recall drop versus OHE, as target-encoding smoothing obscured the sharp signals of niche job roles. These results establish an accuracy ceiling for 4-class classification under categorical mapping, motivating the extended analysis in Section~\ref{subsec:extresults}.

\subsection{Feature Importance and Economic Determinants}
\label{subsec:shap}

\begin{figure}[t]
    \centering
    \includegraphics[width=\linewidth]{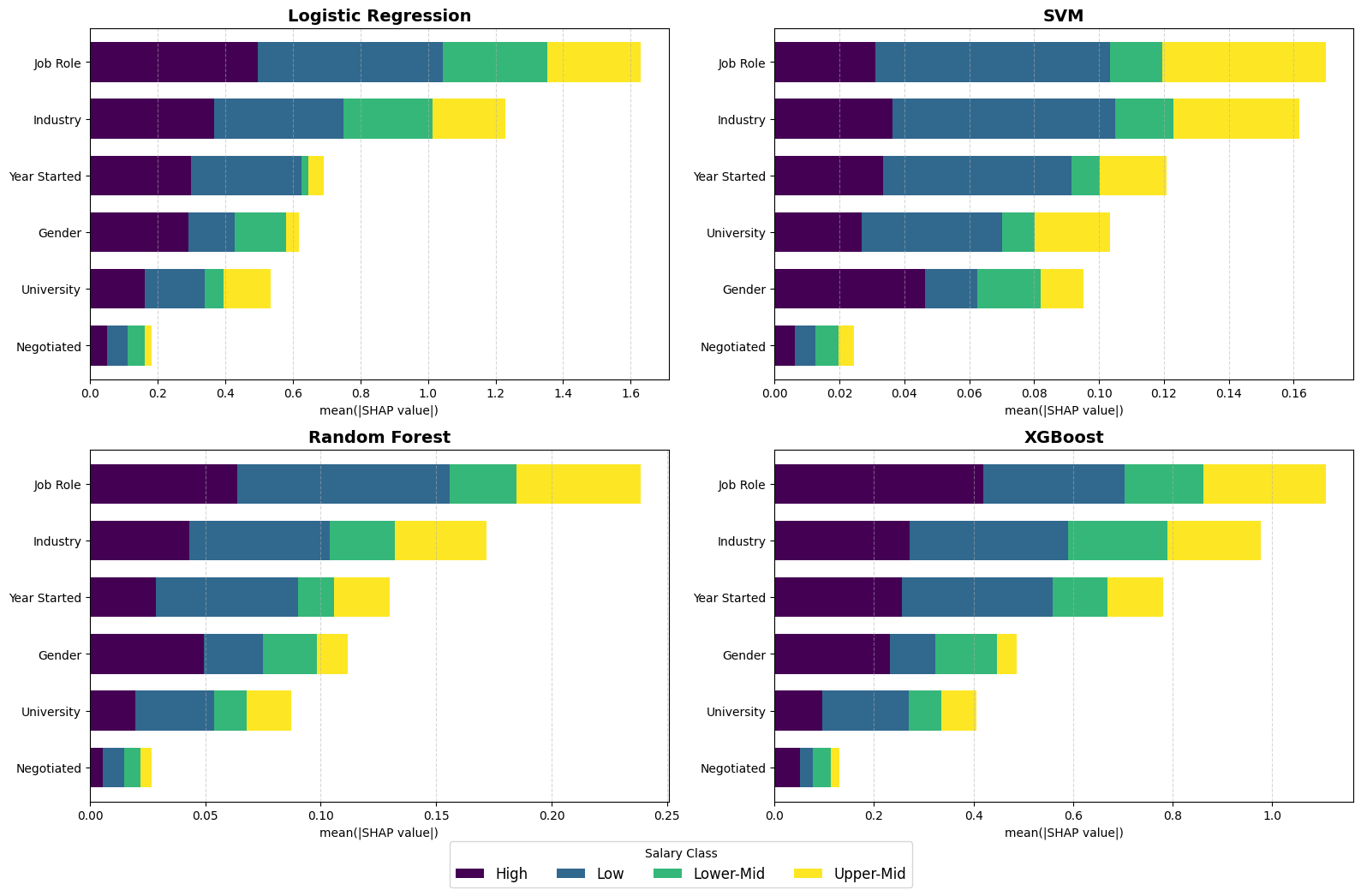}
    \caption{Multi-model SHAP Summary Plots}
    \label{fig:mm_shap}
\end{figure}

Figure \ref{fig:mm_shap} shows mean absolute SHAP values across all four classifiers. All models identify Job Role and Industry as the dominant drivers of starting salary, outweighing University Category by approximately 2$\times$ in the best-performing model. This is the first line of evidence that the local labor market rewards functional skills and sector over institutional pedigree.

Year Started ranks consistently among the highest-importance features, confirming a tenure–compensation relationship. Gender maintains a moderate but consistent influence, which may signal structural pay disparities and warrants further, dedicated investigation rather than causal interpretation from feature attribution alone. Negotiation Status is the least influential feature, consistent with structurally determined entry-level pay in this demographic.

\subsection{Extended Results: Exploratory Diagnosis of the Ceiling}
\label{subsec:extresults}
The classical baseline establishes the paper's central explainable finding. The experiments in this section are exploratory: they probe whether the 43.1\% ceiling reflects limitations in feature representation, model architecture, or task formulation.

\begin{figure}[t]
    \centering
    \includegraphics[width=0.7\linewidth]{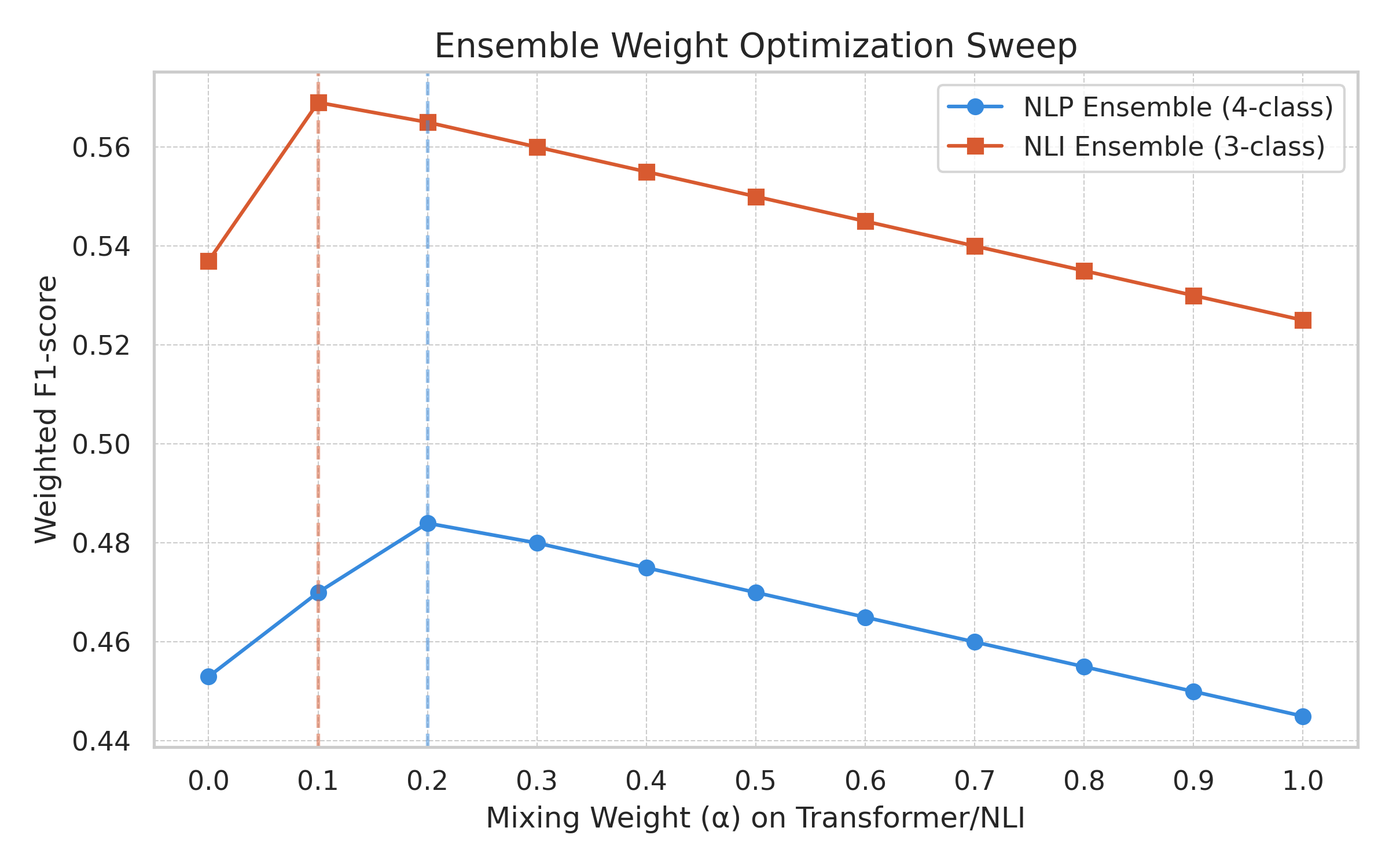}
    \caption{Ensemble weight optimization sweep. In both formulations the best blend leans on the keyword model (LR+TF-IDF), with the optimal transformer/NLI weight $\alpha$ small (0.2 for the 4-class classifier, 0.1 for the 3-class NLI reformulation).}
    \label{fig:alpha_sweep}
\end{figure}

\subsubsection{Performance under NLP Extensions}
Textual representation significantly improved predictive performance on the original 4-class task (Table \ref{tab:nlp_results}). TF-IDF with bigrams yielded a 6.3-point Weighted F1-score improvement over the baseline, confirming that manually mapped categories discarded salary-relevant textual specificity. XLM-RoBERTa fine-tuning further improved performance, though the marginal gain over TF-IDF was modest.

\begin{table}[t]
\centering
\caption{Model performance improvements on the original 4-class task.}
\label{tab:nlp_results}
\renewcommand{\arraystretch}{1.2}
\begin{tabular*}{\textwidth}{@{\extracolsep{\fill}}lccc}
\toprule
\textbf{Method} & \textbf{F1w} & \textbf{Acc} & \textbf{vs. Baseline} \\ \midrule
LR + OHE (baseline) & 0.390 & 43.1\% & --- \\
LR + TF-IDF & 0.453 & 45.4\% & +6.3 pts \\
XLM-RoBERTa fine-tuned & 0.479 & 47.7\% & +8.9 pts \\
\textbf{XLM-RoBERTa + LR ensemble ($\alpha=0.2$)} & \textbf{0.484} & \textbf{49.1\%} & \textbf{+10.1 pts} \\ \bottomrule
\end{tabular*}
\end{table}

The optimal ensemble weight of $\alpha=0.2$ indicates that exact keyword matching accounts for approximately 80\% of the learnable salary signal. This constitutes a second, independent line of evidence for the SHAP finding (Section~\ref{subsec:shap}): the model relies overwhelmingly on the role and industry text, exactly the features SHAP ranks highest. Two methodologically distinct approaches, game-theoretic feature attribution and validation-tuned ensemble weighting, thus converge on the same conclusion about what drives salary.

\subsubsection{Impact of Task Reformulation}
The label reformulation experiments produced the largest performance changes relative to the 4-class baseline (Table \ref{tab:reformulation_results}). The 3-class economic formulation improved F1w by 16.7 points over the baseline, and the binary formulation reached 71.5\% accuracy. These gains, however, are partly a consequence of reduced task granularity: each coarsening of the label space discards predictive resolution that a career-advising tool might need (e.g., the binary split cannot distinguish a near-minimum-wage salary from one just below the median). The fairest cross-task comparison is therefore the margin above each task's random-guess baseline (25\% for 4-class, 33.3\% for 3-class, and 50\% for binary). On this basis the 4-class ensemble ($+24.1$ points) and 3-class NLI ($+23.7$ points) are comparable to, and slightly exceed, the binary result ($+21.5$ points). The reformulation thus relieves, but does not eliminate, the underlying ceiling.
 
This residual ceiling is best understood as a property of the data rather than the models. The Mid salary class is the hardest to predict in every formulation because many different roles and industries map to the same mid-range pay, and the features that would separate them, such as company size, seniority, location, and experience, are absent from the survey. The economic boundaries help by aligning the classes with real labor-market tiers, but they cannot supply information the dataset never contained.

\begin{table}[t]
\centering
\caption{Performance across alternative task formulations.}
\label{tab:reformulation_results}
\renewcommand{\arraystretch}{1.2}
\begin{tabular*}{\textwidth}{@{\extracolsep{\fill}}llccc}
\toprule
\textbf{Method} & \textbf{Task} & \textbf{F1w} & \textbf{Acc} & \textbf{vs. Baseline} \\ \midrule
xlmr + LR, 3-class economic & 3-class & 0.557 & 55.6\% & +16.7 pts \\
\textbf{NLI prompting + LR ($\alpha=0.1$)} & 3-class & \textbf{0.569} & \textbf{57.0\%} & \textbf{+17.9 pts} \\
xlmr + LR, binary median & binary & 0.708 & 70.8\% & +31.8 pts \\
\textbf{Multi-boundary NLI ensemble} & binary & \textbf{0.715} & \textbf{71.5\%} & \textbf{+32.5 pts} \\ \bottomrule
\end{tabular*}
\end{table}

\subsubsection{NLI Corroboration}
The NLI reformulation reached a 3-class F1w of 0.569, the best of any 3-class model. Figure \ref{fig:alpha_sweep} shows how each ensemble splits its weight between the two parts. For the 4-class task, the best blend put about 80\% of the weight on the keyword model (LR+TF-IDF) and 20\% on the transformer. For the 3-class NLI task, the best blend leaned even further on the keyword model, about 90\%, with only 10\% on the NLI model. Across both formulations, then, the role and industry keywords carry most of the predictive power, and adding a reasoning-based model on top changes little. This is the third independent line of evidence for the salary signal: even a model built to reason about the economic \textit{meaning} of a role cannot outweigh the plain role and industry text, echoing what SHAP and the ensemble weight already showed. The NLI component still helps as a small addition, lifting the 3-class result to its best, which suggests the occupational text carries genuine economic structure rather than dataset-specific noise.

\section{Conclusion}
Across all experimental configurations, job role and industry are the primary determinants of starting salary for Filipino graduates, significantly outweighing institutional prestige. This conclusion is unusually well-supported because it emerges from three methodologically independent lines of evidence: game-theoretic SHAP attributions on the classical baseline, the validation-tuned ensemble weight that places most predictive reliance on occupational text, and the NLI reformulation, where the optimal blend still leans about 90\% on the role and industry text, so that even a model built to reason about the economic meaning of a role cannot displace it. This triangulation is what lets the study make an explainable claim rather than merely a predictive one. The accuracy ceiling we observed reflects task design and data quality rather than model capacity. The original quartile boundary falls at the densest salary region, and the residual ceiling is largely a property of a noisy, self-reported survey whose free-text predictors and missing confounders (company size, region, experience) leave the dense mid-salary range hard to separate.
 
For future work, this study establishes two transferable baselines, namely that raw occupational text outperforms standard classification codes and that economically motivated boundaries outperform statistical quartiles, which together suggest career guidance should prioritize sector and role alignment. Future efforts should incorporate the missing confounders, collect data through more controlled instruments, and explore NLP-based role clustering. By bridging descriptive tracer studies and predictive analytics, this work provides a data-driven foundation for understanding what drives entry-level pay in the Philippine labor market.

\bibliographystyle{splncs04}
\bibliography{references}

\end{document}